# Highly curved reflective W-shape and J-shape photonic hook induced by light interaction with partially coated microfluidic channels

Lingya Yu[1], Zengbo Wang[2,*], Lizhou Xu[1], Liyang Yue[2], Bing Yan[2], Igor V. Minin[3], Oleg V. Minin[3], Baidong Wu[2], Yiduo Chen[2]

[1] ZJU-Hangzhou Global Scientific and Technological Innovation Centre, Zhejiang University, Hangzhou, 31120, China.

[2] School of Computer Science and Electronic Engineering, Bangor University, Dean Street, Bangor, Gwynedd LL57 1UT, UK.

[3] Nondestructive school, Tomsk polytechnic University, 30 Lenin Ave., Tomsk 634050 Russia

Email: *z.wang@bangor.ac.uk

**Abstract**

Photonic hook (PH) is a new type of artificial self-bending beam focused by a dielectric particle-lens with a curved waist smaller than the wavelength, which has the potential to revolutionize mesoscale photonics in many applications, e.g., optical trapping, signal switching, imaging, etc. In this paper, we discover a new mechanism that the highly curved PHs can be realised by the light interaction with the fully or partially metal-coated microchannels. The generated W-shaped and J-shaped PHs have bending angles exceeding 80-degree. Compared to other PH setups, the proposed design has a larger range to flexibly control the bending angle through the coating process and can be easily integrated with the established microfluidic systems.

**Introduction**

Conventional optical instruments have limited imaging resolutions due to optical diffraction limit at about half of the incident light wavelength, λ. Under white light illumination, the imaging resolution is about 200-300 nm, making it difficult to observe nanoscale objects and processes[1]. One solution to achieve super-resolution beyond such limit is by photonic nanojet (PNJ), which is a sub-wavelength, high intensity, and low scattering light focus generated by light propagation through a dielectric mesoscale particle and formed on its shadow side[2–5]. It was proved that the imaging resolution of a traditional optical microscope could reach a super-resolution level of λ/8 by placing the dielectric microspheres above the imaging target[6]. Besides, PNJs are promising in other applications, including micro and nanofabrication, microparticle detection, optical manipulation, etc[3,5,7].

Meanwhile, Minin et al. discovered that PNJ has a novel quality for the formation of artificial self-bending light in the near field[8]. This type of PNJ has a curved hook-like field intensity distribution, and for this reason, it is named photonic hook (PH)[8–10]. In principle,
PH forming mechanism requires asymmetry of the dielectric particle in the manner of geometric shape, lighting, or material optical properties[10–13]. These asymmetric dielectric micro-particles allow the light to have different phase velocities in them during the propagation, then inducing a PH in the shadow direction. Now PH has the potential to revolutionize mesotronics in many applications[14], e.g. signal switching[15], optical trapping[16], etc.

The history of focusing light with mirrors has a long history[17]. In this work, we demonstrate a new mechanism to generate the PHs in reflection mode using a concave mirror realised by the coating of a semi-cylindrical microchannel, which is promising for, for example, the analysis of nanoparticles in microfluidics. A reflected PH can be observed at the positions in the middle of the microfluidic channel. Using a numerically electromagnetic simulation, we investigate the formation of the proposed PHs and quantify their key parameters, e.g., angles of curvature, height of curvature and electric field intensity, etc. It shows that the shapes and curvatures of PHs can be precisely controlled by the coating area inside the channel. Consequently, the formed PH is W-shaped (W-hook) under the condition of a majorly coated channel, meanwhile, the PH gradually deforms into a J-shape (J-hook) as the coating area decreases. This makes it possible to implement a dynamically changing focusing, from W- to J- type. According to this finding, the angles measuring PH curvatures are defined in this study, and the above trend of PH shape change is illustrated by the simulated electric field distributions and the function curves of PH curvatures against coating areas in a detailed manner.

**Model**

The three-dimensional numerical model was built by a commercial finite integral technique software package - CST microwave studio (CST). Figure 1 a) and b) show the 3D and 2D cross-sectional view of the model, respectively. The model mimicked a sealed half-cylinder microfluidic channel with a radius of $r$ and arc angle of 180° on a glass substrate with the refractive index, $n_g$ = 1.50. The curved surface of the channel bottom was coated by a perfect electric conductor (PEC) film with a thickness of 100nm. The entire channel was filled with water whose refractive index, $n_w$ is 1.33. A plane wave with the wavelength of 500 nm irradiated from the top of the model and propagated in the direction from +y to -y after the normalisation of its field intensity to 1. The hexahedral meshing with the cell size of $\lambda/10$ was used for the CST frequency domain solver to ensure accuracy. To fully absorb the waves and avoid back-folding, the boundaries at the x/y-axis ends of the model were defined as a perfectly matched absorbing layer. To mimic the infinite length in the z-axis direction, the tangential magnetic fields and normal electric fluxes are set to zero on both z end surfaces of the 3D model (500nm thick along z-axis). The simulated results are shown in the field intensity of $|E|^2$, which is obtained by squaring the magnitude of the electric field.

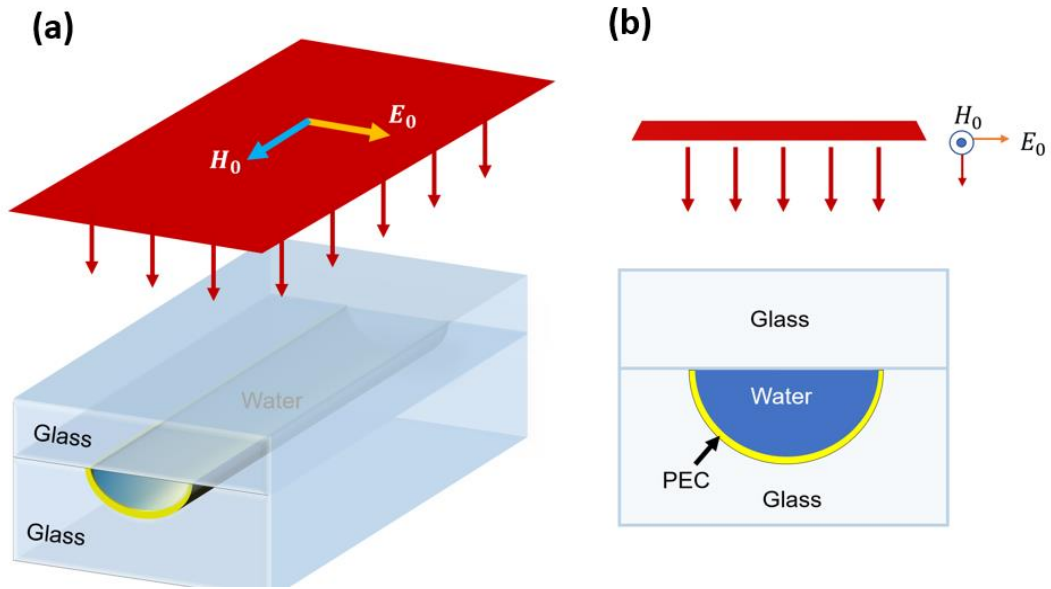

**Figure 1:** Schematic for proposed reflective W-hook and J-hook generation inside a microfluidic channel.

**Results and discussions:**

Firstly, several PHs are demonstrated in Figure 2 by varying the size of the half-cylinder channel under the plane wave illumination. As can be seen, the PEC coating inside the channel can form a concave mirror, which leads to an optical focus in reflection mode. For all the cases, there is a focal point with the strongest intensity located at the centre of the channel, for which the full width at half maximum (FWHM) was measured. The results indicate that the FWHMs at the focal points gradually increase as the growth of channel radius. Accordingly, the FWHM reaches super-resolution with FWHM = 0.288λ at *r*= 5 um channel radius (Fig.2a, this radius corresponding to size parameter $q = 2\pi r/\lambda \approx 62.8$), but increases to 0.705λ -1.1λ for radius larger than 10 um. Figure 2 also demonstrates that the high-intensity curved beams extend to both sides of the channel, then forming a W-shape PH centred at the focal point. In the case of 5 μm radius, the W-shape photonic hook is surrounded by many dot-like spots (Fig.2a). As the channel enlarges, the corresponding intensity of the surrounding dot spots will decrease, resulting in the more pronounced features of PHs. This effect is maximised in the case of the 20-μm-radius channel, as shown in Figure 2 (d), meanwhile, it is analysed as an example to quantify the influence of the coating area on the generated PH. The physical mechanism of W-hook formation can be easily understood in terms of concave mirror caustic aberrations of the 3rd order (spherical and coma)[18,19]. Such an aberrational analysis is applicable for a channel with a radius of more than 10 um, when geometrical optics is valid (size parameter of the mirror is q>100)[3,20].

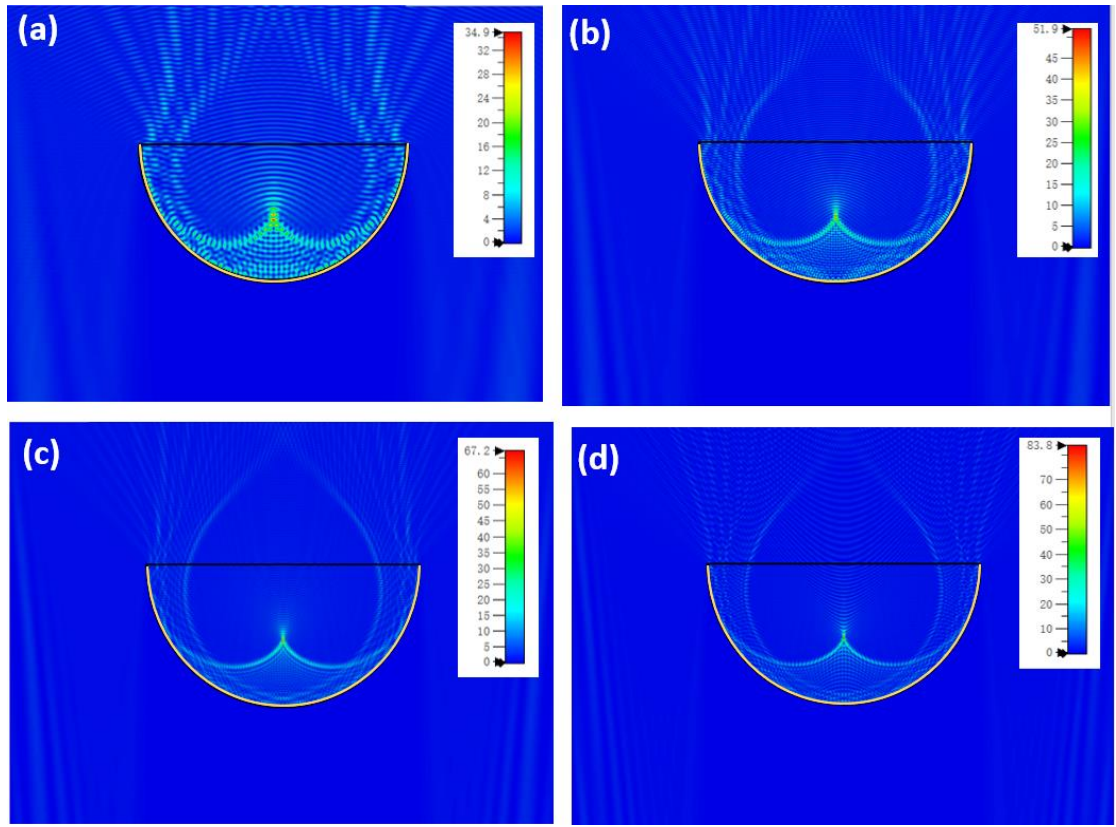

**Figure 2:** Reflective W-shaped photonic hooks, formed inside different sized bottom-coated microfluidic channels, with channel radius (a) 5 µm, (b) 10 µm, (c) 15 µm and (d) 20 µm. Stronger field localizations were formed as size increases.

We now turn to the effect of coating area on the photonic hook. Figures 3 (a)-(h) show the distributions of the PH field intensities when metal coating areas change from 100% to 75%, 58.3%, 50%, 47.22%, 38.89%, 30.58%, and 25% in the 20-µm-radius channel, respectively. The gold line in the figure represents the area covered by the metal coating. A typically symmetric W-hook is shown in Figure 3 (a) for 100% channel coating. As the coating area reduces, the intensity and length of the left side of PH decrease in parallel, leading to a transition of the PH shape from W to J. This phenomenon is caused by the reflected beam with the asymmetric phase velocity and interference led by the partial coating of the channel[11]. As can be seen in Fig. 3 (c), when the area of the concave metal mirror is reduced to 58.3%, the beam on the left side of the PH completely disappears, producing a more uniform intensity distribution on the right side of the PH. Thus, as the coating area further reduces, the curvatures of the J-hooks become smaller, represented as the ends of the PHs (left part) gradually approaching the channel profiles, as shown in Figure 3 (d) to (h). Reducing the coating area increases the off-axis aberrations of the focused beam, which leads to a change in the shape and curvature of the hook. The effective area of the mirror that forms the focused beam also decreases.

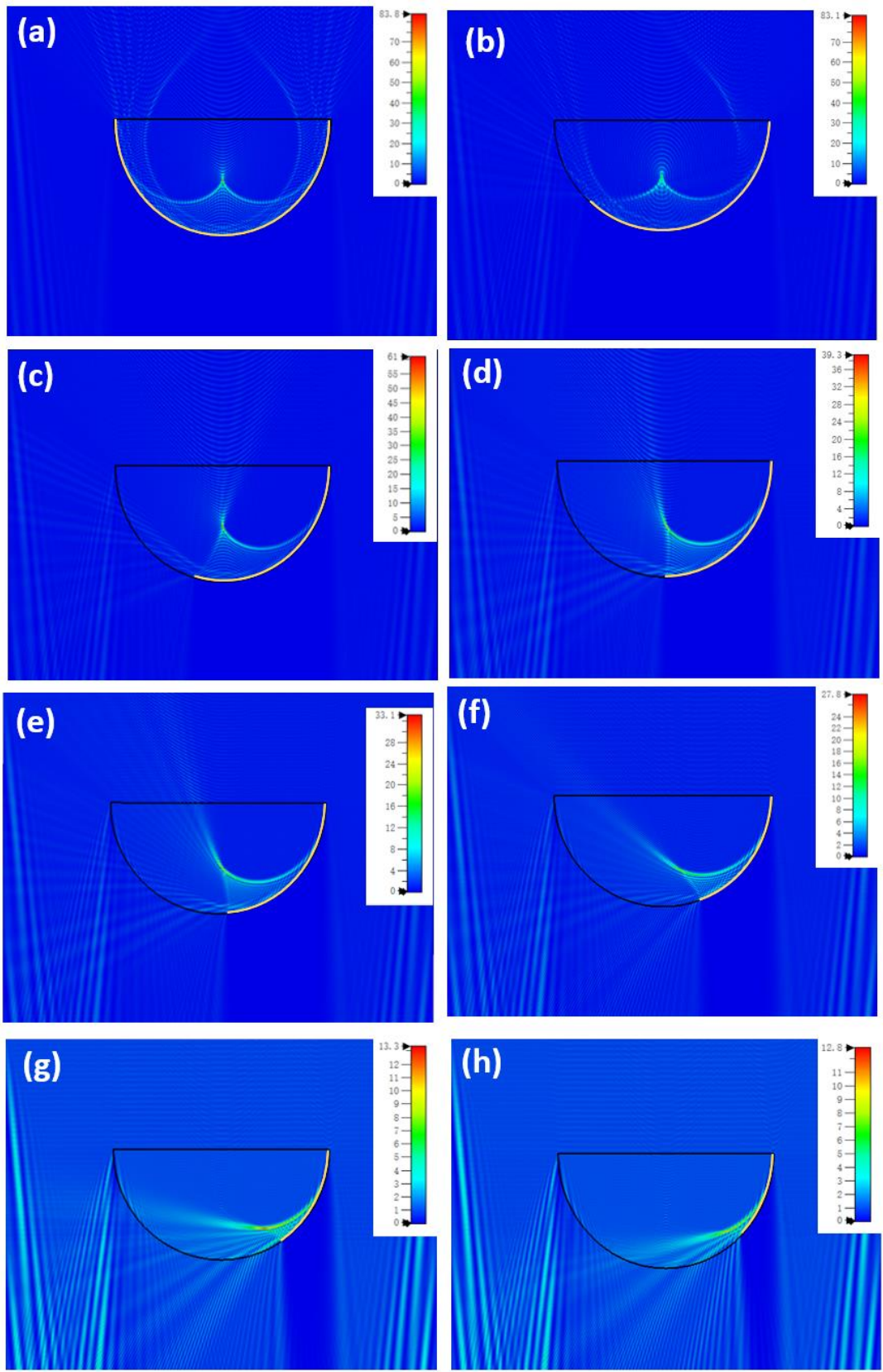

**Figure 3:** Evolution of photonic hook from W-shape to J-shape inside a 20-μm-radius microfluidic channel via coating engineering, with coating area (a) 100%, (b) 75%, (c) 58.33%, (d) 50%, (e) 47.22%, (f) 38.89%, (g) 30.58% and (h) 25% respectively.

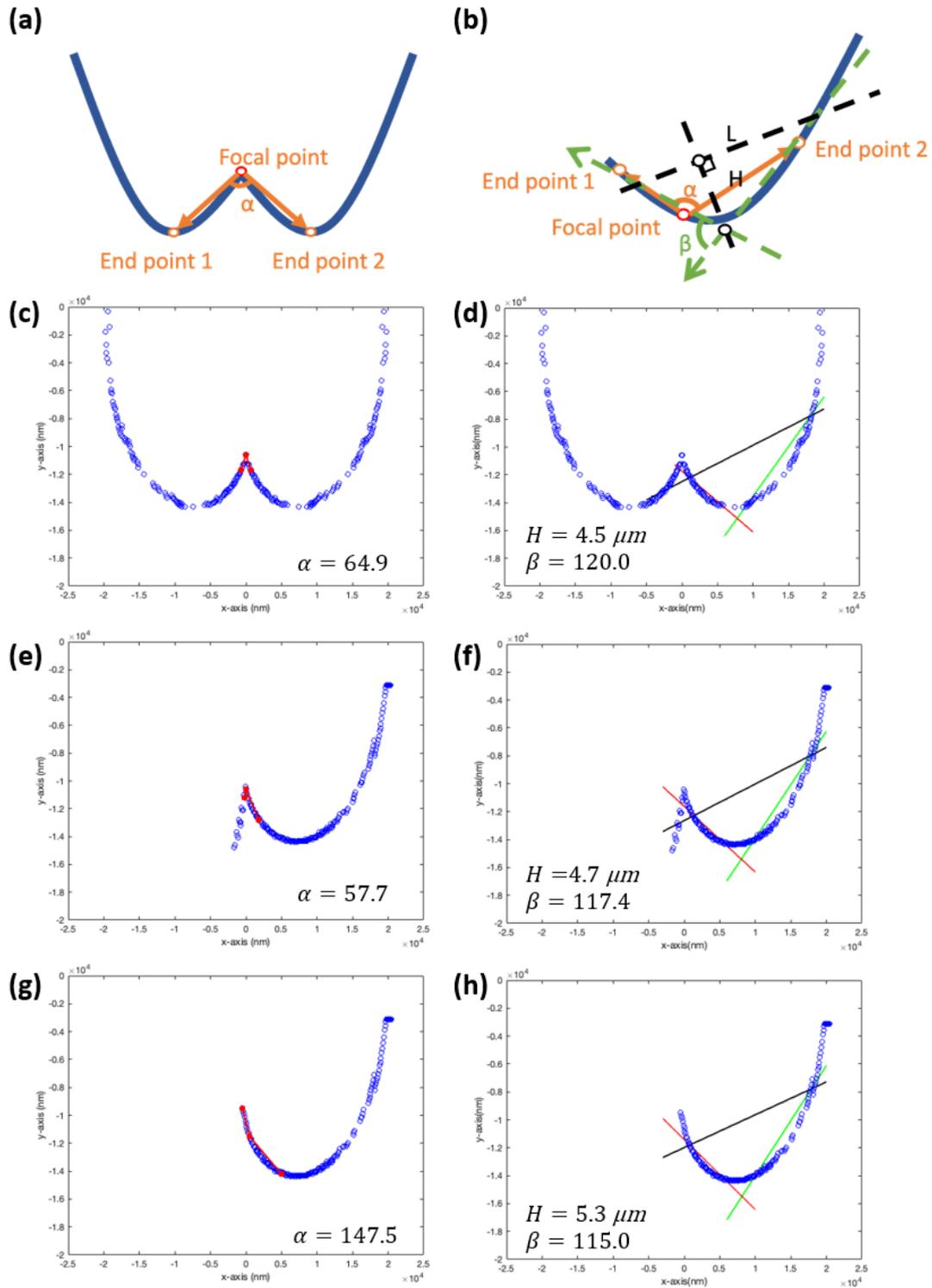

**Figure 4:** Extracted Hook profile and angles as in Figure 3. (a) angle $\alpha$ definition: peak intensity dropping to 1/e, (b) angle $\beta$ definition: intersection angle between left wing and right wing of a J-hook. Index-H definition: the distance between the intersection of two wings and center of subtense. (c) and d) shows the measurement of angle $\alpha$, angle $\beta$ and index-H for 100% coating area. e) and f) 58.33%, g) and h) 50%.

Several parameters were used to characterize the curvature of the localized photonic flux and the variation of the field intensity distribution for PH[12,13]. In this work, the focal point is defined as the point of maximum intensity in a PH for both W-hook and J-hook, as shown in Figure 4 (a) and (b), and their ends are the positions with the field strength reaching 1/e of that maximum intensity. In this case, the angle $\alpha$ is measuring the angle between the PH two ends with the focal point as the apex. The magnitude of angle $\alpha$ is influenced by the PH variation. For this reason, it is noted that the W-hooks normally deliver an angle $\alpha$ opening downwards, as shown in Figure 4 (a). By contrast, because of the asymmetry of J-hooks, their angle $\alpha$ have an upwards opening, as shown in Figure 4 (b). Thus, the angle $\alpha$ is well defined to distinguish the W-hook and J-hook. Figure 4 (b) also shows another curvature parameter - angle β which is a measure for the curvature between the fitted left and right parts of the PH dividing by the bending point in the middle (the smaller angle). The main difference between angle $\alpha$ and angle β is that β is not affected by the intensity of the PH, but simply reflects the change in the shape of the PH. In addition, the measurement of hook curvature is also indicated by the height index H. As shown in Fig.4 b), the index H shows the distance between hook arc centre and the line L, where L is a line connected by two intersections of left and right wings with best-fit lines. Obviously, a larger value of the H-index indicates a larger curvature of the hook. Fig. 4 (c)-(h) illustrate the profiles of the W-hooks and J-hooks for the models with the coating percentages of 100%, 58.33% and 50% through plotting the PH intensities varying the position with a step size of 100 nm, and all angle α, β index-H measurements are indicated in them as well.

Also, the corresponding trends of angle $\alpha$, angle β, and H-index against the coating area are summarized in Fig. 5 (a), (b) and (c), respectively. It is noted that there is a large difference in the angles $\alpha$ for the W-hooks and J-hooks in Figure 5 (a). All $\alpha$ angles of W-hooks are smaller than 80°, but the smallest angle β already reaches the level of 120°. This dramatic change happens at the coating percentage of about 55% which is larger than the half coating threshold of 50%. Besides, the fluctuation of the angle β curve is not as strong as that for angle $\alpha$, and is represented as a decline in J-hook region with coating percentages that are smaller than 50%. This is according to the tendency that the J-hooks approach the channel bottom with the decrease of the coating area, as shown in Figure 2 (d) to (h). The trend of the H-index is essentially similar to that of the angle β, with a rapid upward trend at a coating area of 50% and a peak at a coating of around 40%, followed by a rapid downward trend. The H-index and angle beta parameters reflect the bending property of the photonic hook, and the similarity of the two trends fully validates the bending property of the photonic hook obtained from the model.

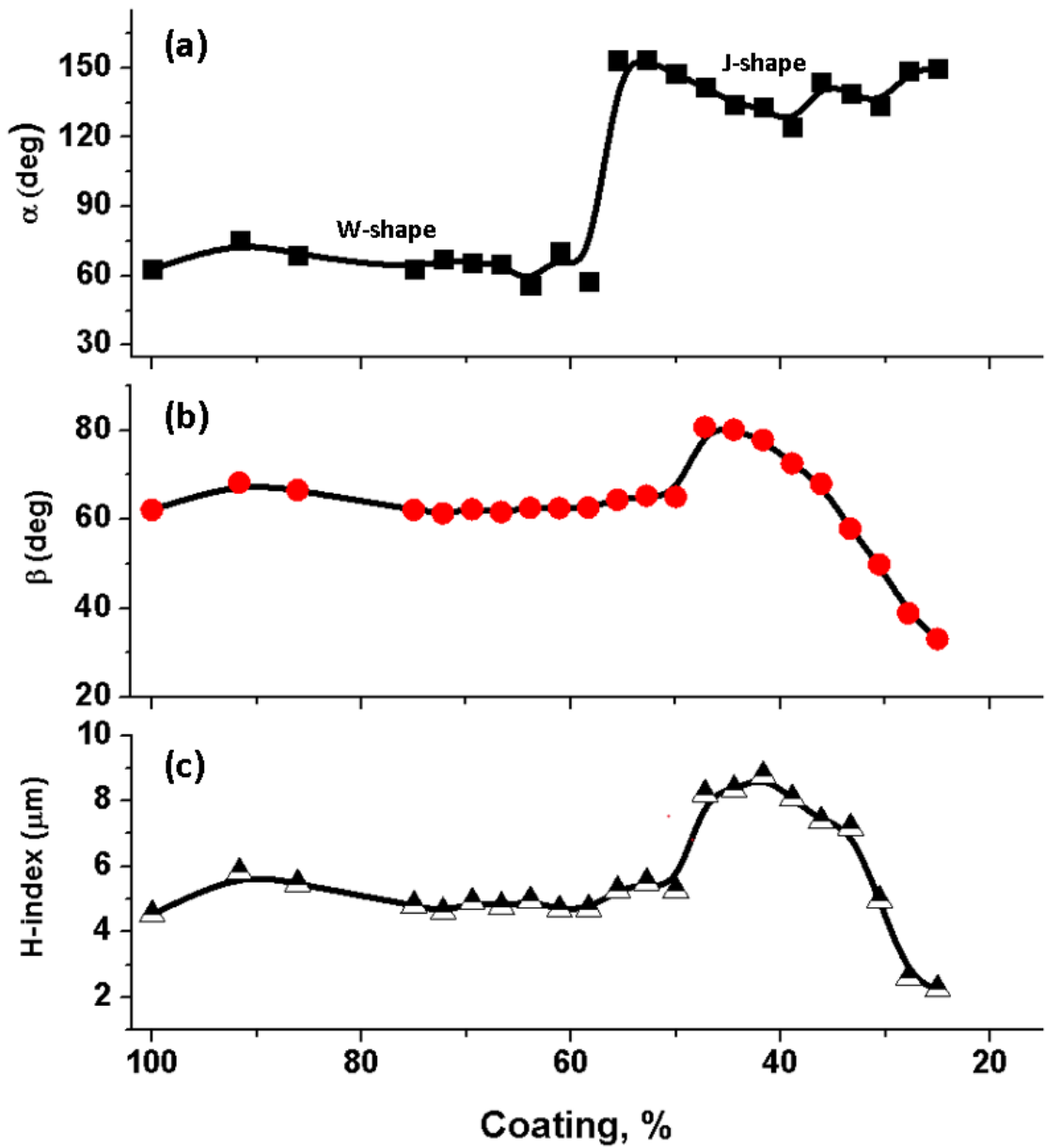

**Figure 5:** W and J Hook angles and height as a function of coating size. (a) $\alpha$-angle, (b) $\beta$-angle and (c) H-height.

**Conclusion**

To summarize, we report a new optical mechanism to generate a self-bending PH by the concave mirror. It is formed by irradiation of the plane wave light to a semi-cylindrical micro-channel whose curved bottom is coated with metal film. By varying the area of the coating, the originally symmetrical reflection and its spatial phase velocity are broken to form two types of PHs in W-shape and J-shape, respectively. The transition between W-hook and J-hook can be accomplished by the adjustment of the coating area, and the turning point is reported as about 55% coating coverage in this study. The coating process can flexibly control the bending angle of the proposed PHs in a large range, and the entire system can be integrated with the established micro-channel systems.

**Acknowledgement.** *This research was supported by the Welsh European Funding Office (WEFO); European ERDF grants, No. CPE 81400 and SPARCII c81133; UK Royal society (IEC\R2\202178 and IEC\NSFC\181378), and partially supported by TPU development program.*

**Data Access Statement:** *The data that support the findings of this study are available from the corresponding author upon request.*

**Conflict of Interest declaration:** *The authors declare no conflict of interest.*